\documentclass[12pt,a4paper]{article}
\pdfoutput=1

\usepackage{jcappub}

\hypersetup{colorlinks,bookmarksopen,bookmarksnumbered,citecolor=blue,
linkcolor=black,pdfstartview=FitH,urlcolor=blue}

\usepackage{graphicx}
\usepackage{amsmath}
\usepackage{amssymb}

\def\bea{\begin{eqnarray}}
\def\eea{\end{eqnarray}}
\def\ba{\begin{eqnarray}}
\def\ea{\end{eqnarray}}
\def\be{\begin{equation}}
\def\ee{\end{equation}}
\def\beq{\begin{equation}}
\def\eeq{\end{equation}}

\newcommand{\lsim}{\mathrel{\rlap{\lower4pt\hbox{\hskip1pt$\sim$}}
    \raise1pt\hbox{$<$}}}         
\newcommand{\gsim}{\mathrel{\rlap{\lower4pt\hbox{\hskip1pt$\sim$}}
    \raise1pt\hbox{$>$}}}         

\newcommand{\leftrightarrowraised}{\mathrel{\rlap{\lower-0pt\hbox{\hskip1pt$\partial$}}
    \raise6 pt\hbox{$\leftrightarrow$}}}

\newcount\hour \newcount\minute
\hour=\time \divide \hour by 60
\minute=\time
\title{A halo-independent lower bound on the dark matter capture rate in the Sun from 
a direct detection signal}

\def\kth{Department of Theoretical physics,
School of Engineering Sciences, KTH Royal Institute of Technology,
AlbaNova University Center, 106 91 Stockholm, Sweden}

\def\okc{Oskar Klein Centre for Cosmoparticle Physics, Department of Physics,
Stockholm University, SE-10691 Stockholm, Sweden}

\author[a]{\textbf{Mattias Blennow,}\vspace*{0mm}}

\author[a]{\textbf{Juan Herrero-Garcia}\vspace*{0mm}}
\affiliation[a]{\kth}

\author[b]{\textbf{and Thomas Schwetz}\vspace*{0mm}}
\affiliation[b]{\okc}

\abstract{We show that a positive signal in a dark matter (DM) direct
  detection experiment can be used to place a lower bound on the DM
  capture rate in the Sun, independent of the DM halo. For a given
  particle physics model and DM mass we obtain a lower bound on the
  capture rate independent of the local DM density, velocity
  distribution, galactic escape velocity, as well as the scattering
  cross section. We illustrate this lower bound on the capture rate by
  assuming that upcoming direct detection experiments will soon obtain
  a significant signal. When comparing the lower bound on the capture
  rate with limits on the high-energy neutrino flux from the Sun from
  neutrino telescopes, we can place upper limits on the branching
  fraction of DM annihilation channels leading to neutrinos. With
  current data from IceCube and Super-Kamiokande non-trivial limits
  can be obtained for spin-dependent interactions and direct
  annihilations into neutrinos. In some cases also annihilations into
  $\tau\tau$ or $b b$ start getting constrained. For
  spin-independent interactions current constraints are weak, but they
  may become interesting for data from future neutrino telescopes.}


\begin{document}
\begin{flushright}
 NORDITA-2015-17
\end{flushright}
\maketitle

\section{Introduction}

From gravitational effects we know that dark matter (DM) constitutes a
significant fraction of the energy density of the Universe. Among the
most promising ways to search for non-gravitational manifestations of
DM particles are direct detection (DD) experiments, which are looking
for the scattering of DM particles from the galactic halo in
underground detectors~\cite{Bernabei:2010mq, Ahmed:2011gh,
  Aprile:2012nq, Archambault:2012pm, Akerib:2013tjd, Angloher:2014myn,
  Agnese:2013rvf, Agnese:2014aze}, and neutrino telescopes looking for
high-energy neutrinos from the annihilations of DM particles in the
Sun~\cite{Desai:2004pq, Tanaka:2011uf, Aartsen:2012kia,
  Adrian-Martinez:2013ayv, Choi:2015ara}. The latter signal emerges from the capture
of galactic DM particles in the gravitational potential of the Sun
after loosing enough energy in a DM--nucleus scattering event in the
Sun. Hence, the capture rate of DM in the Sun is determined by the
scattering cross section of the DM particle on nuclei, the same
process which provides the signal in DD experiments. The resulting
neutrino flux from the Sun will also depend on the annihilation
channels of the DM particle. Therefore, viable information on DM
properties can in principle be obtained by comparing the two signals.

However, the DD signal and the DM capture in the Sun
depend on different parts of the DM velocity distribution. While DD
experiments are sensitive to DM particles with velocity larger than a
certain minimal velocity (which depends on the mass of the nuclei in the detector and its energy threshold as well as on the DM mass), the DM capture in the Sun is sensitive to values below a certain maximum velocity, above which capture of DM
particles is kinematically forbidden. Therefore,
in order to explore the complementarity of the two signals, it is common to adopt specific DM velocity distributions, for instance the
so-called Standard Halo Model (SHM) consisting of a truncated
Maxwellian distribution, see refs.~\cite{Kamionkowski:1994dp,
  Bergstrom:1998xh, Ullio:2001yf, Hooper:2008cf, Wikstrom:2009kw,
  Kappl:2011kz, Arina:2013jya, Liang:2013dsa} for a very incomplete
list of examples of this approach. However, the properties of the DM
velocity distribution as well as the local DM density are plagued with
large uncertainties and halo-independent methods are desirable to draw
robust conclusions about possible signals. The impact of variations of
halo properties on the neutrino signal have been studied for instance
in refs.~\cite{Bruch:2009rp, Choi:2013eda} and in the context of the
neutrino/DD comparison in ref.~\cite{Serpico:2010ae}. The authors of
ref.~\cite{Kavanagh:2014kx} investigated the potential to extract DM
parameters from a combination of data from DD experiments and from a
neutrino signal based on a polynomial parameterization of the DM
velocity distribution, whose parameters are fitted together with the
DM parameters.

In the present paper we develop a completely halo-independent
method to compare a signal from a DD experiment with a neutrino signal
from the Sun. We show that from a precise measurement of the DD nuclear
recoil spectrum a halo-independent lower bound on the capture rate in
the Sun can be derived. It is based on the overlap region in velocity
space and therefore does not require any assumptions about the halo
properties. Our bound extends the halo-independent methods developed
in the context of DD \cite{Fox:2010bz, Fox:2010bu} to the capture rate
in the Sun.

The remainder of this paper is organized as follows: We review the
phenomenology of DD in section~\ref{s:D} and of the DM induced
neutrino signal from the Sun in section~\ref{s:ID}. We then discuss
the relation of direct detection to the capture rate in the Sun in
section~\ref{s:DID}, where the central result of the paper (the lower
bound on the capture rate) is given in section~\ref{bound}. In
section~\ref{numerics}, we apply the bound to mock data from future
direct detection experiments and compare them to the upper bounds from
the IceCube and Super-Kamiokande neutrino telescopes. We
  also comment on the importance of nuclear form factor uncertainties
  as well as on the ratio of the neutron and proton couplings. We summarize and
give our concluding remarks in section~\ref{conc}.  In
appendix~\ref{ap:modulations}, we discuss how to use an annual
modulation signal in a DD experiment to provide the lower bound on the
capture rate, and apply our results to the DAMA signal. In
appendix~\ref{app:v} we show how to use our results in the case of
more general scattering cross sections, beyond contact interactions.

\section{Dark matter direct detection} \label{s:D}

In this section we review the relevant expressions for DD of dark
matter \cite{Goodman:1984dc}. We focus on elastic scattering of DM
particles $\chi$ with mass $m_\chi$ off a nucleus with mass number $A$
and mass $m_A$, depositing the nuclear recoil energy $E_R$. The
differential rate (measured in events/keV/kg/day) for a single target
detector is:\footnote{For detectors with several nuclei the total rate
  is the sum of the rates in all nuclei, i.e., $\mathcal{R}(E_R, t) =
  \sum_A \mathcal{R}_A(E_R, t)$, but usually one nucleus gives the dominant contribution to the
  rate for a particular DM mass.}
\begin{equation}\label{eq:R}
  \mathcal{R}(E_R, t) = \frac{\rho_\chi}{m_\chi m_A}
  \int_{|\vec{v}| > v_m} d^3 v  \, v f_{\rm det}(\vec{v}, t) \frac{d \sigma_A}{d E_R}(v)  \,,
\end{equation}
with $\rho_\chi$ being the local DM mass density and $v_m$ is the minimal
velocity of the DM particle required for a recoil energy $E_{R}$:
\beq \label{vmin}
v_{m}=\sqrt{ \frac{m_A E_{R}}{2 \mu_{\chi A}^2}} \,,
\eeq
where $\mu_{\chi A}$ is the reduced mass of the DM--nucleus system.
The function $f_{\rm det}(\vec{v}, t)$ describes the distribution of
DM particle velocities in the detector rest frame, with the normalization 
\beq\label{eq:etaf1} 
\int  d^3 v \, f_{\rm det}(\vec{v}, t) = 
\int_0^\infty  d v\, v^2 \tilde{f}_{\rm det} (v,t) = 1 \,,
\eeq
and we define the angular averaged velocity distribution function
$\tilde{f}$ by
\begin{equation}\label{eq:ftilde}
\tilde f(v) \equiv \int d\Omega f(v, \Omega) \,,
\end{equation}
where $d\Omega =d\cos\theta \,d\phi$. The velocity
distributions in the rest frames of the detector, the Sun and the
galaxy are related by 
$
f_{\rm det}(\vec{v},t) = f_{\rm Sun}(\vec{v} +
\vec{v}_e(t))=f_{\rm gal}(\vec{v} + \vec{v}_s+\vec{v}_e(t)) \,, 
$
where $\vec{v}_e(t)$ is the velocity vector of the Earth relative to
the Sun and $\vec{v}_s$ is the velocity of the Sun relative to the
galactic frame. The revolution of the Earth around the Sun encoded in
$\vec{v}_e(t)$ leads to an annual modulation of the DD signal
\cite{Drukier:1986tm, Freese:1987wu}.

To be specific, in the following we will concentrate on
spin-independent (SI) and spin-dependent (SD) scattering from a
contact interaction. This implies that the differential scattering
cross section $d\sigma_A(v)/dE_R$ scales as $1/v^2$. Our results can
also be generalized to other $v$ dependences, as shown in
appendix~\ref{app:v}. For SI contact interactions with equal DM couplings to
neutrons and protons the cross section becomes
\begin{equation}\label{eq:CS}
  \frac{d\sigma_A}{dE_R}(v) = \frac{m_A \sigma_\chi^p A^2}{2\mu^2_{\chi p} v^2} F^2_A(E_R) \,,
\end{equation}
where $\sigma_\chi^p$ is the total DM--proton scattering cross
section at zero momentum transfer, $\mu_{\chi p}$ is the DM--proton
reduced mass, and $F_{A}(E_R)$ is a nuclear form factor.  For SD
interactions a similar formula applies with no $A^2$ enhancement and a different form factor.

Then the event rate eq.~\eqref{eq:R} becomes
\begin{align}  
\mathcal{R}(E_R, t) = \,A^2  F_{A}^2(E_R) \, \tilde\eta (v_m, t) \,,
\label{eq:R0}
\end{align}
where $v_m$ is considered as a function of $E_R$ according to eq.~\eqref{vmin} 
and we have defined
\beq\label{eq:eta} 
\tilde{\eta}(v_m, t) \equiv \mathcal{C}\,\eta(v_m, t) \quad\text{with}\quad \eta(v_m, t) \equiv 
\int_{v_m}^\infty  d v \,v \tilde{f}_{\rm det} (v,t)
\quad\text{and}\quad
\mathcal{C} \equiv  \frac{\rho_\chi \sigma_\chi^p }{2 m_\chi\mu_{\chi p}^2} \,.
\eeq 
Notice that $\tilde \eta$ depends only on the DM properties and the DM
halo, but is independent of the characteristics of the experiment.
For fixed DM mass, one can translate the event rate in $E_R$ space
into $v_m$ space, and $\tilde\eta(v_m, t)$ then has to be the same for
any experiment. This is the basis of the halo-independent methods
developed in refs.~\cite{Fox:2010bz, Fox:2010bu} and used extensively to
compare results of different DD experiments, see, e.g.,
refs.~\cite{McCabe:2011sr, McCabe:2010zh, Frandsen:2011gi,
  HerreroGarcia:2011aa, HerreroGarcia:2012fu, DelNobile:2013cta,
  DelNobile:2013cva, Bozorgnia:2013hsa, Fox:2014kua, Feldstein:2014ufa, Cherry:2014wia, Bozorgnia:2014gsa}.

For a specific detector the number of DM induced events in an energy range 
between $E_1$ and $E_2$ is given by 
\beq \label{Nevents} 
N_{[E_1,E_2]} = M T A^2 \int_0^\infty d E_R \, F_{A}^2(E_R) G_{[E_1, E_2]}(E_R) 
\tilde{\eta} (v_m, t) \,,
\eeq 
where $M$ and $T$ are the detector mass and exposure time
respectively, and $G_{[E_1,E_2]}(E_R)$ is the detector response
function describing the probability that a DM event with true recoil
energy $E_R$ is reconstructed in the observed energy interval
$[E_1,E_2]$, including energy resolution, energy dependent
efficiencies, and possibly also quenching factors.

In section~\ref{bound} we will assume that a positive signal is observed in a direct
detection experiment. In this case, the angular averaged distribution $\tilde f(v)$
times the constant $\mathcal{C}$ can be extracted from the data
(modulo experimental resolutions and uncertainties). Using
eqs.~\eqref{eq:R0} and~\eqref{eq:eta}, we find \cite{Drees:2007hr}
\begin{equation}\label{eq:deriv}
\mathcal{C} \tilde f(v) =  -\frac{1}{v}\frac{d \tilde\eta(v)}{d v} 
=  - \frac{1}{v A^2} \frac{d}{d v} \left(\frac{\mathcal{R}(E_R)}{F_A^2(E_R)}\right),
\end{equation}
where $E_R$ is considered as a function of $v = v_m$ according to
eq.~\eqref{vmin}, depending on the DM mass.

\section{The neutrino signal from DM annihilations in the Sun} \label{s:ID}

In this section we briefly review the relevant expressions for the
neutrino signal from the Sun. Due to the scattering of DM particles
with the nuclei in the Sun they may lose energy and become
gravitationally bound to the Sun. Their annihilation products can
produce neutrinos with energies comparable to the DM mass, detectable
at Earth \cite{Press:1985ug,
  Griest:1986yu,Gould:1987ju}.

\begin{figure}
	\centering
	\includegraphics[width=0.45\textwidth]{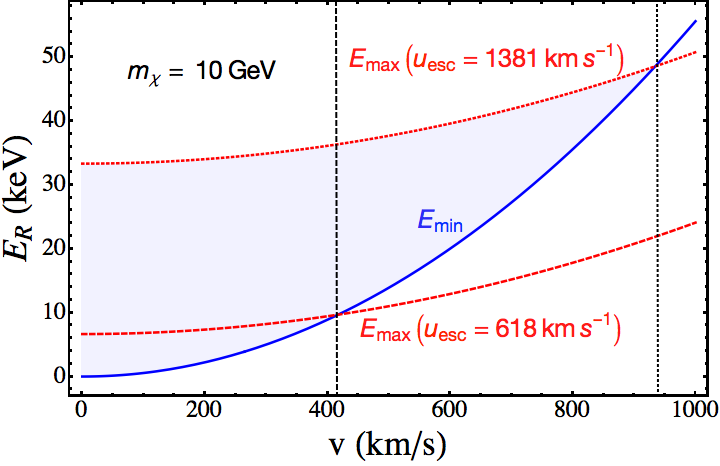}~~
	\includegraphics[width=0.45\textwidth]{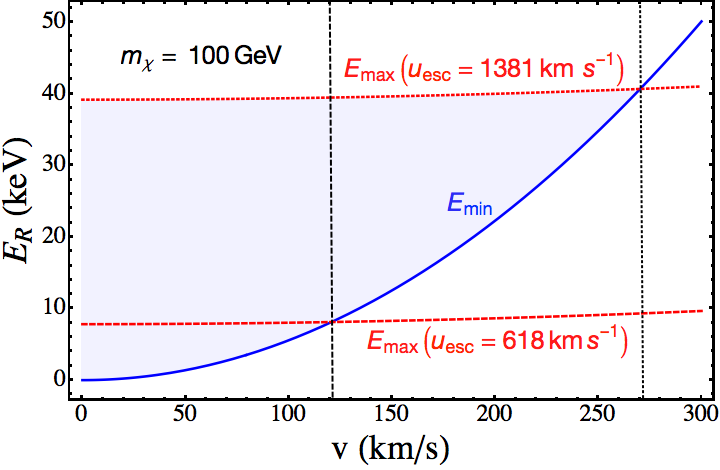}
	\caption{For hydrogen (SD), we show in blue the lower bound on the energy for DM capture in the Sun, $E_{\rm min}$, and in red the energy upper limits, $E_{\rm max}(r)$, for the two extreme escape velocities, versus the velocity v, for $m_\chi=10 \, (100)$ GeV, left (right). The points were they cross (indicated by the vertical dotted lines) are the maximum velocities of the DM particles, $v_{\rm cross}^p (r)$.} \label{Erv}
\end{figure}

The capture rate of dark matter particles is given by 
(see, e.g., refs.~\cite{Gould:1987ju, Peter:2009mk})
\beq  \label{capt}
C_{\rm Sun}= 4\pi \frac{\rho_\chi}{m_\chi} 
\sum_A \int_0^{R_{\rm Sun}} dr\, r^2 \int_0^{\infty} dv \tilde{f}(v) \,v\, w \,
\Omega_A(w, r) \,,
\eeq 
where, for SI interactions, the sum over $A$ goes over all element abundances
in the Sun up to nickel, while for SD interactions only hydrogen is relevant. We use the short-hand notation $\tilde{f}_{\rm Sun } \equiv \tilde{f}$ and the angular averaged
velocity distribution function is defined in
eq.~\eqref{eq:ftilde}. The function $\tilde f(v)$ refers to the velocity
distribution at infinity, whereas $w$ is the DM velocity inside the
gravitational potential of the Sun at the radial distance $r$ from the
solar center, given by $w^2 = v^2 + u_{\rm esc}^2(r)$, with $u_{\rm
  esc}^2(r)$ being the escape velocity from the Sun depending on the
location $r$.  The quantity $\Omega_A(w, r)$ is the rate with which a DM particle
with velocity $w$ will be gravitationally captured by scattering on
the nucleus $A$ in a spherical shell at radius $r$ with thickness
$dr$:
\begin{equation} \label{Omega}
  \Omega_A(w,r) = w \, \frac{\rho_A}{m_A} \int_{E_{\rm min}(w)}^{E_{\rm max}(w)} dE_R 
  \frac{d\sigma_A}{dE_R}(w) \,,
\end{equation}
where $\rho_A(r)$ is the mass density of the element $A$ in the Sun
($\rho_A/m_A$ being the number density).  For the numerical
calculations we use the standard solar model from
ref.~\cite{Serenelli:2009yc}. Here, $E_R$ is the 
recoil energy of the nucleus after the scattering. 
For a given DM velocity $w$ there is a minimal and
maximal nuclear recoil energy, $E_{\rm min}(w)$ and $E_{\rm max}(w)$,
such that the DM particle gets trapped, i.e., its velocity after the
scattering is less than the local escape velocity. Writing them in
terms of the velocity $v$ at infinity one has
\beq \label{Em}
E_{\rm min} = \frac{m_\chi}{2}v^2 \,,\qquad 
E_{\rm max} = \frac{2 \mu_{\chi A}^2}{m_A} (v^2+u^2_{\rm esc}(r)) \,,
\eeq
where the former constraint is the requirement for the DM particle to be captured and the latter is based on the maximal energy transfer allowed by kinematics.
In figure~\ref{Erv} we show $E_{\rm min}$ and $E_{\rm max}(r)$ versus
velocity $v$ for scattering on hydrogen for two different DM masses,
$m_\chi=10$ and 100~GeV, and for the two extreme values of the escape
velocity in the Sun: $618\, {\rm km \,s^{-1}}$ at the surface and
$1381\, {\rm km \,s^{-1}}$ in the center of the Sun. One can see that
$E_{\rm min}$ and $E_{\rm max}(r)$ cross, which defines the maximum
velocity the DM particles can have in order to be trapped in a single interaction, $v^A_{\rm cross}(r)$. From
eq.~\eqref{Em} we find
\beq \label{vcross}
v_{\rm cross}^A (r) =\frac{\sqrt{4 m_A m_\chi}}{|m_\chi-m_A|}u_{\rm esc} (r)
\qquad\text{or}\qquad 
E_{\rm cross}^A(r)=\frac{2 m_A m_\chi^2}{(m_\chi-m_A)^2} u^2_{\rm esc} (r) \,.
\eeq
Therefore, $v_{\rm cross}(r)$ decreases with DM mass $m_\chi$, 
and the capture rates will decrease and eventually vanish for large DM masses.
For SD interactions, where only hydrogen is relevant, we can use $m_p \ll m_\chi$ and 
\beq
v_{\rm cross}^p (r) \approx 2 \sqrt{\frac{m_p}{m_\chi}} u_{\rm esc} (r) 
\qquad \text{or}\qquad 
E_{\rm cross}^p  \approx2 m_p u_{\rm esc} (r) \,.
\eeq
Typically, for heavier nuclei $v_{\rm cross}^A (r)$, which is relevant for
SI interactions, is larger than $v_{\rm cross}^p (r)$, relevant for SD interactions.  

The cross section $d\sigma_A(w)/dE_R$ in eq.~\eqref{Omega} is the same
as the one entering in the event rate for DD
experiments. Again, restricting to contact interactions\footnote{For
  other types of interactions see appendix~\ref{app:v}.}, we use
eq.~\eqref{eq:CS} for SI interactions and an analogous expression for
SD. For the nuclear form factors of the elements in the Sun we use
the approximation $F_A^2(E_R)\simeq e^{-E_R/E_A}$, with
$E_A=3/(2m_AR_A^2)$ and $R_A = [0.91(m_A/{\rm GeV})^{1/3} +
  0.3]$~fm~\cite{Gould:1987ju}. Hence, we obtain
\beq \label{capture}
C_{\rm Sun}= 4\pi \, \mathcal{C}\sum_A A^2 \int_0^{R_{\rm Sun}} dr r^2 \rho_A(r) \int_0^{v_{\rm cross}^A} dv \tilde{f}(v)\,v\, \int_{E_{\rm min}(v)}^{E_{\rm max}(v)} F_A^2(E_R) dE_R 
\eeq 
for the capture rate in SI interactions.
The coefficient $\mathcal{C}$ is defined in eq.~\eqref{eq:eta} and
contains the DM--nucleus scattering cross section.  For SD
interactions only hydrogen is relevant, i.e., the sum contains only
one term with $A=1$, and the form factor is trivial, $F_{\rm H}^2(E_R)
= 1$.

If equilibrium between DM capture and annihilation in the Sun is
reached, the final annihilation rate is independent of the
annihilation cross section and is given by: 
\beq \label{Gsun} 
\Gamma_{\rm Sun}=\frac{1}{2} C_{\rm
  Sun} \,.  
\eeq 
For equilibrum to occur, the equilibration time $t_{\rm eq}$ must be smaller than the age of the Sun, i.e., $t_{\rm eq}\ll t_{\rm Sun}\sim 4.5\,\rm Gyr$, where (see for instance ref.~\cite{Kappl:2011kz})
\beq \label{teq} 
t_{\rm eq}=\frac{1}{\sqrt{C_{\rm Sun}\,A_{\rm Sun}}}\approx 0.5\, {\rm Gyr} \,
\Big(\frac{10^{21}\,\rm s^{-1}}{C_{\rm Sun}}\Big)^{1/2}\,\Big(\frac{3\cdot10^{-26}\,\rm cm^3\,s^{-1}}{\langle \sigma v\rangle}\Big)^{1/2}\, \Big(\frac{100\,\rm GeV}{m_\chi}\Big)^{3/4} \,.  
\eeq 
Here $A_{\rm Sun}$ is the annihilation rate in the Sun and $\langle
\sigma v\rangle$ the thermal average of the annihilation cross
section. As we will see in section~\ref{numerics} for mock data and in
appendix~\ref{ap:modulations} for DAMA, our values of the lower bound
on the capture are safely above $\gtrsim10^{21}\,\rm s^{-1}$ in all
cases except for SI interactions in xenon, due to the small scattering
cross section assumed for the mock data, $\sigma_{\rm SI}=10^{-45}\,
\rm cm^3\,s^{-1}$. In this case, equilibrium may not be reached for
annihilation cross sections smaller or equal than the freeze-out
one. Also we note that in case of $p$-wave annihilations the
annihilation cross section today can be much smaller than the thermal
freeze-out value assumed in eq.~\eqref{teq} and depending on the
capture rate equilibrium may or may not be reached.

Notice that we are neglecting evaporation, which is justified for $m_\chi\gtrsim 3 $~GeV
\cite{Griest:1986yu, Gould:1987ju, Busoni:2013ek}. The neutrino flux at the Earth from the annihilation channel $f$ with branching ratio BR$_f$ is
\beq
\frac{d\phi_\nu^f}{dE_\nu}={\rm BR}_f \frac {\Gamma_{\rm Sun}}{4\pi d^2} \frac{dN_\nu^f}{dE_\nu} \,,
\eeq
where $d$ is the Sun-Earth distance and $dN_\nu^f/dE_\nu$ is the neutrino spectrum per annihilation of flavour $f$ that reaches a distance of 1~AU, which needs to take into
account flavour transitions (neutrino oscillations, MSW effect),
absorption and regeneration, see, e.g., refs.~\cite{Blennow:2007tw,
  Cirelli:2008lq}.

\section{Relating DM direct detection to the DM capture rate in the Sun} 
\label{s:DID}

We now want to relate a DD signal to the DM capture
rate in the Sun in a halo-independent way. We assume that
the DM velocity distributions relevant in the two cases are the
same. This implies two important consequences:
\begin{enumerate}
\item 
We neglect the small velocity of the Earth with respect to the Sun 
and adopt the approximation for $\tilde{f}$ (defined in eq.~\eqref{eq:ftilde})
\[
\tilde f_{\rm det}(v) \approx \tilde f_{\rm Sun}(v) \equiv \tilde f(v) \,.
\]
Hence we use that $v_e \approx 29\,{\rm km/s} \ll v_m$ for typical
values of $v_m$.  This  also implies that we are ignoring the (small)
annual modulation signal in the direct detection rate (we will comment on this in appendix~\ref{ap:modulations}).

\item
DM direct detection samples the DM distribution today, while for the
DM capture in the Sun the velocity distribution on time scales
relevant for the equilibration of capture and annihilations is
relevant. We will assume that the DM velocity distribution is constant
on those time scales and that the same $\tilde f(v)$ applies for the
neutrino capture and direct detection. Similarly, we also assume that
the energy density is constant on times scales relevant for
equilibration and equal to the current one.
\end{enumerate}

\subsection{The overlap in $v_m$}

From inspection of the relations given in sec.~\ref{s:D} we see that
direct detection is sensitive to \emph{high} DM velocities. Let us denote the threshold energy of a given experiment by $E_{\rm thr}$. If we ignore
the finite energy resolution of the experiment, this defines a
threshold velocity via eq.~\eqref{vmin}: $v_{\rm thr} \equiv v_m
(E_{\rm thr})$.\footnote{In the case of a finite energy resolution the
  actual threshold velocity would be somewhat smaller than the value
  corresponding to the nominal threshold energy of the experiment due
  to the reconstruction of events below the threshold at higher
  energies.}  Hence, this experiment is sensitive to DM velocities $v
> v_{\rm thr}$. In a realistic halo there will be a maximal velocity
of DM particles, set by the escape velocity from the galactic
gravitational potential, $v_{\rm esc}$. Since the precise value of
this escape velocity is uncertain (see, e.g., ref.~\cite{Lavalle:2014rsa}
and references therein) we want to keep the discussion as
independent of this argument as possible.

In contrast to direct detection, eq.~\eqref{capture} shows that only DM velocities $v <
v_{\rm cross}^A(r)$ contribute to the capture rate in the Sun. Hence, direct detection is sensitive to DM velocities above a certain value given by $v_{\rm thr}$, whereas for DM to be captured in the Sun velocities below the value set by $v_{\rm cross}^A(r)$ are
relevant. In order to relate the two phenomena we need to consider the
overlap regions in velocity space. If $v_{\rm cross}^A < v_{\rm thr}$,
there is no overlap and direct detection and DM capture \emph{decouple}. In this case no
statement can be made halo-independently and a connection can be
established only by referring to some a-priori assumptions about the DM
velocity distribution.

\begin{figure}
  \centering
  \includegraphics[width=0.5\textwidth]{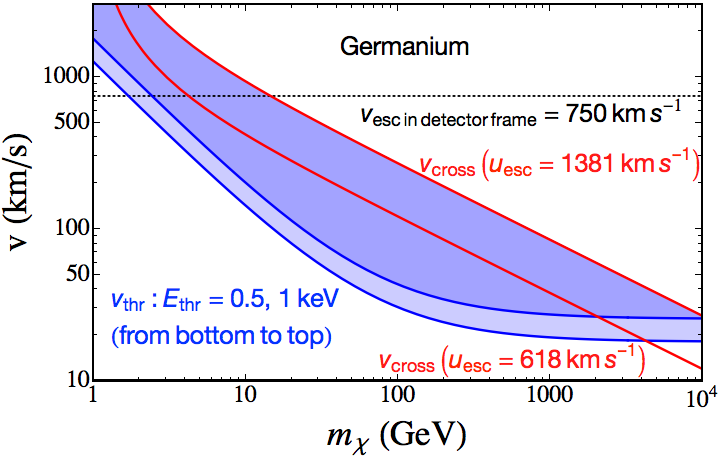}~	
  \includegraphics[width=0.5\textwidth]{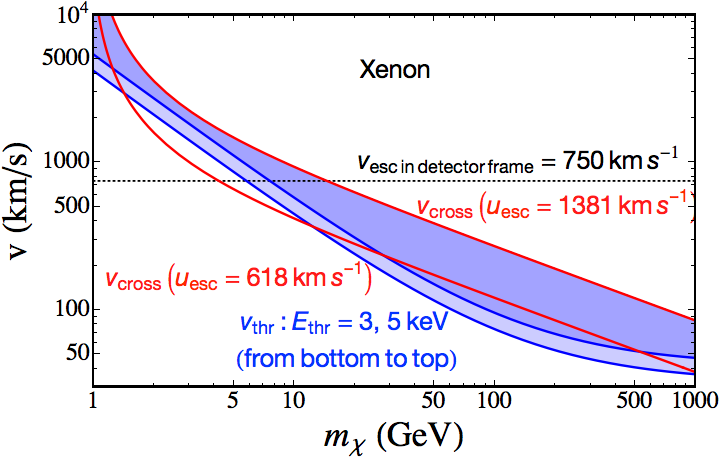}
  \caption{In blue we show the minimum velocity $v_{\rm thr}(E_{\rm
      thr})$ probed in a direct detection experiment versus $m_\chi$,
    assuming different threshold energies $E_{\rm thr}$ and using a Ge
    (left) and a Xe (right) target. In red we show the maximum
    velocity relevant for DM capture in the Sun for scattering on
    hydrogen, $v_{\rm cross}^p$, for the two extreme values of $u_{\rm
      esc} = 1381 \, {\rm km \, s^{-1}}$ in the centre of the Sun and
    $v_{\rm esc} = 618 \, {\rm km \, s^{-1}}$ at the surface. The
    shaded area shows the overlap region assuming scattering in the
    centre of the Sun. The horizontal black line indicates
    approximately the galactic escape velocity in the detector rest
    frame.} \label{DID}
\end{figure}

In figure~\ref{DID} we show with blue curves the minimum velocity
probed in direct detection experiments versus the DM mass, using Ge (left)
and Xe (right) for different examples of threshold energies. Those
curves can be compared with the maximal velocity relevant for the DM
capture in the Sun, $v_{\rm cross}$, which is shown with red curves for scattering on
hydrogen for the two extreme values of the solar
escape velocity $u_{\rm esc}$ corresponding to the values in the
centre and at the surface of the Sun. From the figure we observe that
for a large range of DM masses there is an overlap of the DM velocities
probed by direct detection and solar capture, i.e., $v_{\rm thr} <
v_{\rm cross}$, at least for scattering events which occur near
the centre of the Sun. A similar figure can be found in
ref.~\cite{Kavanagh:2014kx}, where it is also shown that 
$v_{\rm cross}$ is even larger for heavier nuclei
relevant for SI interactions and therefore the overlap 
is also enhanced with respect to SD interactions. 

In the following we are going to make use of the overlap region in
velocity space in order to compare results from direct detection
experiments to neutrino searches from the Sun without specifying any
halo properties apart from the two assumptions stated at the beginning
of this section. In particular, in section~\ref{bound} below we are
going to assume a positive signal in direct detection and derive a
halo-independent lower bound on the capture rate in the Sun. We will use the fact that from a precise measurement of the
  nuclear recoil spectrum the angular averaged distribution $\tilde
  f(v)$ times the constant $\mathcal{C}$ can be extracted from the
  data via eq.~\eqref{eq:deriv}, and we can make use of the signal
  precisely in the overlap region in velocity space.

Before we proceed with this, let us briefly comment on the case of a
positive signal from neutrinos.  In this case no direct information
can be obtained about $\tilde f(v)$, and the signal can come from any
DM velocity below $v_{\rm cross}$. For a positive neutrino signal no
general lower bound for the direct detection rate can be established
halo-independently, since the capture may happen entirely from
velocities below the threshold of the DD experiment. If no signal is
seen in direct detection at the relevant level, such a situation could
point to a halo dominated by low velocity DM particles (in the solar
frame), for instance a co-rotating dark disk
\cite{Bruch:2009rp}. Alternatively this may indicate more exotic
particle physics such as self-interacting DM, which enhances the
capture rate by DM--DM scattering, independently of the DM--nucleus
scattering cross section \cite{Zentner:2009is}. Another example of
a modified relation between the DD signal and capture in the Sun is
inelastic DM scattering~\cite{Nussinov:2009ft, Menon:2009qj,
  Shu:2010ta}. These cases will be studied in a future work.

\subsection{A lower bound on the capture rate from a positive direct detection signal} 
\label{bound}

We can derive a lower bound on the capture rate, eq.~\eqref{capture},
by using that for any positive function $H(v)\geq0$ we have that
$\int_0^{v_{\rm cross}^A} H(v)\,dv \geq \int_{v_{\rm thr}}^{v_{\rm
    cross}^A} \,H(v)\, dv$ for $v_{\rm cross}\geq v_{\rm thr}$. Therefore we have
\beq  \label{boundgeneral}
C_{\rm Sun}
\geq 4\pi \sum_A A^2\mathcal{C} \int_0^{R_{\rm Sun}} dr r^2 \rho_A(r) \int_{v_{\rm thr}}^{v_{\rm cross}^A} dv\, \tilde{f}(v)\, v  \, \mathcal{F}_A(v,r),
\eeq
where we defined
\beq \label{Fdef}
\mathcal{F}_A(v,r) \equiv \int_{E_{\rm min}(v)}^{E_{\rm max}(v)} F_A^2(E_R) dE_R \,,
\eeq
with $E_{\rm min}(v)$ and $E_{\rm max}(v)$ given by eq.~\eqref{Em}. Notice that $E_{\rm max}(v)$, 
$\mathcal{F}_A(v,r)$ and $v^A_{\rm cross}$, eq.~\eqref{vcross}, depend on $r$ via $u_{\rm esc}(r)$.

Let us assume now that a significant direct detection signal has been
observed. We can then use eq.~\eqref{eq:deriv} to obtain $\tilde f(v)$
from the data and insert it in eq.~\eqref{boundgeneral}:
\beq  \label{SI}
\begin{aligned}
C_{\rm Sun}&\geq 4\pi \sum_A A^2 \int_0^{R_{\rm Sun}} dr r^2 \rho_A(r) 
\int_{v_{\rm thr}}^{v_{\rm cross}^A} dv\, \left( - \frac{d \tilde\eta(v)}{d v}\right)\, \mathcal{F}_A(v,r)\\
&=4\pi \sum_A A^2 \int_0^{R_{\rm Sun}} dr r^2 \rho_A(r)\,
\left[ \tilde{\eta}(v_{\rm thr}) \mathcal{F}_A(v_{\rm thr}, r) +  
\int_{v_{\rm thr}}^{v_{\rm cross}^A} dv \,\tilde{\eta}(v)\,\mathcal{F}'_A(v,r)\right] \,,
\end{aligned}
\eeq
where in the last line we integrated by parts, with
$\mathcal{F}_A(v^A_{\rm cross},r) = 0$ and $\mathcal{F}'_A(v,r) \equiv d
\mathcal{F}_A(v,r) / dv$. 
For SD interactions only hydrogen is relevant and we have $F_{\rm H}(E_R)=1$, and 
\beq 
\mathcal{F}_{\rm H}(v,r) = E_{\rm max}(v, r) - E_{\rm min}(v) \,,\qquad
\mathcal{F}'_{\rm H}(v,r) = \left(\frac{4 \mu_A^2}{m_A} -m_\chi \right) v \leq 0 \,.
\eeq 
The direct detection signal has to be
precise enough for $\tilde \eta(v)$ to be extracted from the
observed energy spectrum via eq.~\eqref{eq:deriv} with sufficient
precision. Either the derivative or the function $\tilde \eta(v)$,
including its value at the experimental threshold, have to be determined,
including unfolding of experimental resolutions and backgrounds.  This
will require a significant number of events such that an accurate
spectral analysis can be performed.  If those conditions are met,
eq.~\eqref{SI} provides a lower bound on the capture rate in the Sun
without specifying the DM velocity distribution, the galactic escape
velocity, the scattering cross section or the local DM density.
This lower bound is the central result of this paper and we will illustrate it
numerically  in the following section for possible future signals in DD experiments.

\section{Numerical examples} \label{numerics}

\subsection{Mock data for direct detection}
\label{mock}

The halo-independent bound on the DM capture rate in the Sun derived
in the previous section will be useful once a clear and highly
significant signal in a direct detection experiment has been
observed.\footnote{The CDMS collaboration reports 3
    candidate events from their data with a silicon target, with a
    0.19\% probability for the known-background-only hypothesis when
    tested against the alternative DM+background hypothesis
    \cite{Agnese:2013rvf}. In order to apply our bounds a detailed
    spectral measurement is required and not enough information can be
    extracted from the 3 observed events in CDMS-Si. However,
    motivated by this potential signal, we take DM parameter
    values similar to those preferred by CDMS-Si events for our example of mock
    data in germanium. The annual modulation signal reported by the
    DAMA/LIBRA collaboration~\cite{Bernabei:2010mq} will be discuss in
    appendix~\ref{ap:modulations}. We emphasize that both signals are
    either excluded or in strong tension with several other
    experiments~\cite{Ahmed:2011gh, Aprile:2012nq, Archambault:2012pm,
      Akerib:2013tjd, Angloher:2014myn, Agnese:2014aze} (see for
    instance refs.~\cite{HerreroGarcia:2012fu, Bozorgnia:2014gsa} for
    halo-independent analyses).} In the following we will assume that
the DM--nucleon scattering cross section is just below the current
limits \cite{Agnese:2014aze, Akerib:2013tjd}, which will allow
upcoming experiments to obtain a significant signal. As representative
examples we will consider a future xenon based experiment
\cite{Malling:2011va,Baudis:2012bc,Aprile:2012zx} as well as a
germanium detector \cite{Brink:2012zza, supercdms-snolab}. For
calculating mock data we adopt the conventional Maxwellian velocity
distribution (SHM) with $\bar v = 220$~km/s, truncated at the escape
velocity of $v_\mathrm{esc} = 544$~km/s, and we assume a local DM
density $\rho_\chi = 0.3$~GeV/cm$^3$. The velocity of the Sun in
galactic coordinates is $(10,\, 233,\, 7)$~km/s.

For the xenon experiment we adopt a threshold of 5~keV (we also show the effect of a reduced threshold of 3~keV below) and we
take natural abundances of the isotopes with spin, $^{129}$Xe
($26.4\,\%$) and $^{131}$Xe ($21.2\,\%$).  We use the values of
$10^{-45}\, \rm cm^2$ for the SI cross section and $2\cdot 10^{-40}\,
\rm cm^2$ for the SD cross section, with equal couplings to protons
and neutrons in both cases.  Assuming $m_\chi=100$ GeV, for an exposure of
1~ton~yr at 100\% efficiency
about $154$ ($267$) events would be observed in the energy range $5-45$~keV for SI (SD)
case. 

We also generate mock data for a future germanium experiment, with a
threshold of $1$~keV, focusing on low DM masses. Assuming a DM mass
$m_\chi=6$~GeV and cross sections of $\sigma_{\rm SI}=5\cdot
10^{-42}\, \rm cm^2$ and $\sigma_{\rm SD}=2\cdot 10^{-40}\, \rm cm^2$
(with equal couplings to protons and neutrons) we would obtain about
$1.5\times 10^4$ (2--3) events for SI (SD) interactions in the energy
range 1--10~keV for an exposure of 100~kg~yr with energy resolution of
30\%. While the SI case will allow for a high statistics
reconstruction of the event spectrum, this will not be possible for
the SD case with the assumed parameters. This follows from the missing
$A^2$ enhancement with respect to SI interactions, as well as the small abundance of only 7\% of the
relevant Ge isotope with spin. Nevertheless we are going to include
SD interactions in the following discussion, keeping always in mind that
this will require larger exposures or a different target nucleus with
better sensitivity to SD interactions. In addition, the bounds are stronger for larger DM masses, but we want to provide an illustrative example for the small mass region.

In the following analysis we neglect the energy resolution, effects of
binning the data, possible contamination with background, experimental
errors, and nuclear form factor uncertainties (we 
  comment on form factors later in subsection~\ref{uncertainties}).  Hence, we assume that
  $\tilde\eta(v_m)$ can be extracted from the observed nuclear recoil
  spectrum (for a specified $m_\chi$). This idealized analysis
  suffices to illustrate the power of our bounds. Once applied to real
  data an appropriate statistical analysis will have to be performed.

\begin{figure}[t]
	\centering
        \includegraphics[width=0.47\textwidth]{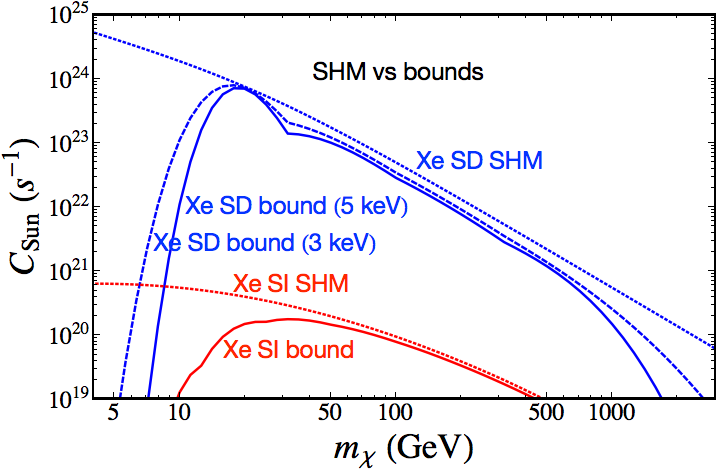}~~  \includegraphics[width=0.5\textwidth]{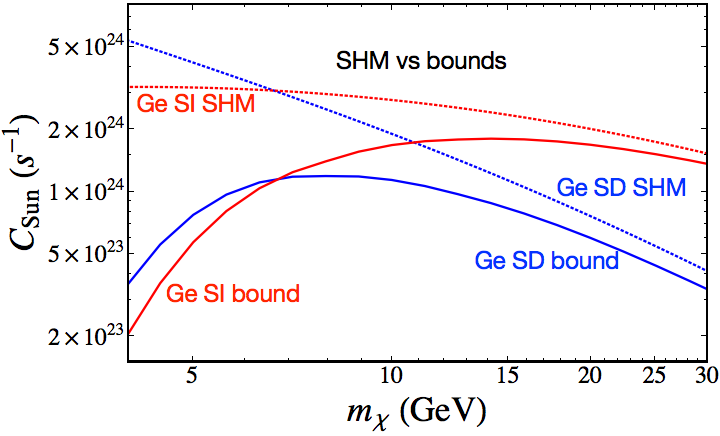}
	\caption{Lower bounds on the DM capture rate in the Sun for
          xenon (left) and germanium (right) DD mock data as described in the text, both for SI
          (red curves) and SD (blue curves) interactions. For illustrative purposes, in the Xe SD
          case we show the effect of changing the default threshold
          energy of $E_{\rm thr} = 5$~keV (solid) to 3~keV (dashed). The dotted curves
          correspond to the actual capture rate. To calculate the DD
          mock data as well as the capture rate we assume the standard
          halo model.} \label{boundsvscap}
\end{figure}

In figure~\ref{boundsvscap} we show the lower bound on the capture
rate for the mock data of future DD xenon (left) and germanium (right)
experiments as described above, both for SD and SI interactions and
compare them to the true capture rate assuming the SHM. In the case of
Ge we focus on the low DM mass region, i.e., $m_\chi \lesssim 30$ GeV,
keeping in mind that the bounds are stronger (i.e., closer to the true
capture rate) for $m_\chi
\gtrsim20$~GeV. The bounds as well as the capture rate are shown for
the ``true'' DM mass, which has been used to calculate the mock
data. We see that our bounds are strong in a large portion of parameter space for the xenon
experiment: $20\lesssim m_\chi\lesssim1000$~GeV for SD, and for
$m_\chi \gtrsim50$~GeV for SI.

\subsection{Limits from neutrino telescopes}

In order to compare the lower limit on the DM capture rate in the Sun
from a DM direct detection signal with the upper bounds on the
neutrino rate from neutrino telescopes we proceed as follows. We
assume equilibrium between capture and annihilations and the
neutrino--induced muon rate in neutrino telescopes will then depend on
the specific annihilation channel, which depends on the particle
physics model for DM. In particular cases, the high-energy neutrino
rate may be strongly suppressed by the available annihilation
channels, for instance for annihilations into $e^\pm, \mu^\pm$ or
$u,d, s$ quarks (see, however, ref.~\cite{Ibarra:2014vya} for higher
order effects). Hence, translating a given upper bound on the
neutrino--induced muon rate into a bound on the capture rate depends
on the DM annihilation channel, see section~\ref{s:ID}.

Below we are going to compare the lower bounds on $C_{\rm Sun}$ to the
limits from IceCube (IC) and Super-Kamiokande (SK). For IC we use the
upper bounds of ref.~\cite{Aartsen:2012kia}, where results are given
directly as upper limit on the capture rate for various annihilation
channels as a function of the DM mass. We will show results for two
cases, namely annihilations into $b b + \tau \tau$ and into
$WW + \tau \tau$ (we keep particle/antiparticle notation implicit for annihilation products). For SK there are no limits on the capture
rate available. Therefore, for annihilation
channels into $b b$ and $\tau \tau$ in the low DM mass region (up to $200$~GeV) we extract the upper limits on the capture from the upper limits on the scattering cross section given by the most recent SK results~\cite{Choi:2015ara}. For the large mass region in these channels, and also for direct annihilation into neutrinos ($\nu_\mu \nu_\mu$), we use the limits on the capture rate calculated in ref.~\cite{Guo:2013ypa} (Tab.~II) based on SK data
from ref.~\cite{Tanaka:2011uf}.
The quoted bounds at the $90\%$~CL from IC
and SK apply assuming that annihilations proceed with 100\% branching
ratio into the indicated channels.

\subsection{Comparison of direct detection and neutrino data}

In order to apply the lower bound on the capture rate derived in
section~\ref{bound} one has to specify the DM mass and the couplings to neutrons and protons. For the moment we will restrict the analysis to equal couplings to neutrons and protons, and we will come back to this point in section~\ref{uncertainties}. Regarding the DM mass, in general it cannot be extracted from a DD signal without
referring to a specific DM halo model. Therefore, we do the analysis
without assuming that the DM mass is known. We calculate mock data for
a fixed ``true'' DM mass, but then we apply the lower bound on $C_{\rm
  Sun}$ as a function of $m_\chi$ (different from the ``true''
value). This procedure resembles the situation we would face in case
of applying the method to real data. If the mass was known from some
other data (e.g., an observation at LHC or a $\gamma$ line signal from
indirect DM searches) one would of course perform the analysis only for
that DM mass.

\begin{figure}[t]
	\centering
        \includegraphics[width=0.5\textwidth]{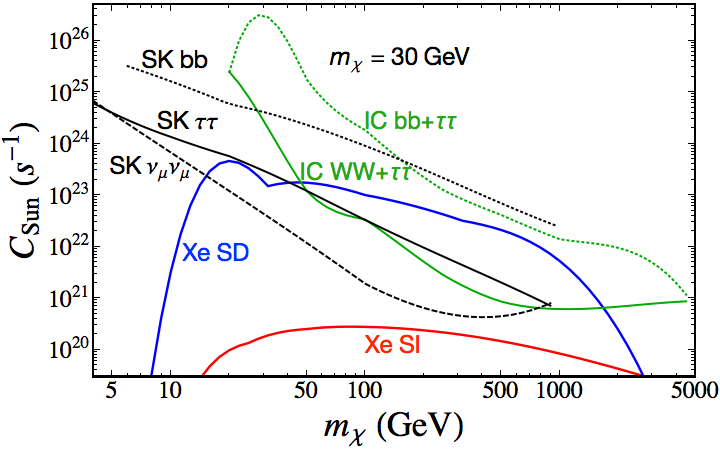}~\includegraphics[width=0.49\textwidth]{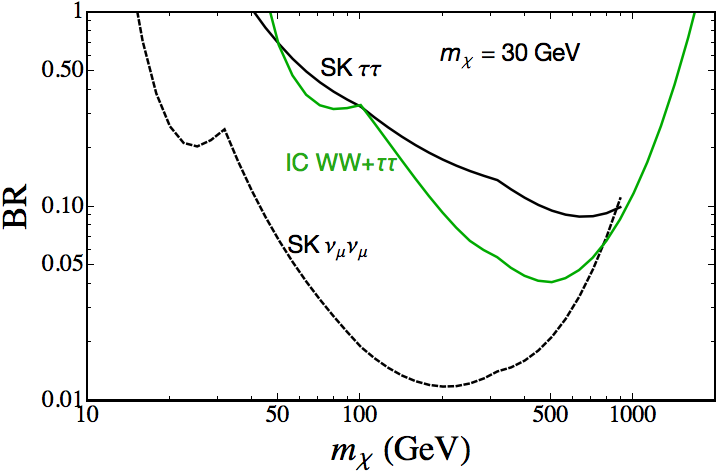}\\
        \includegraphics[width=0.5\textwidth]{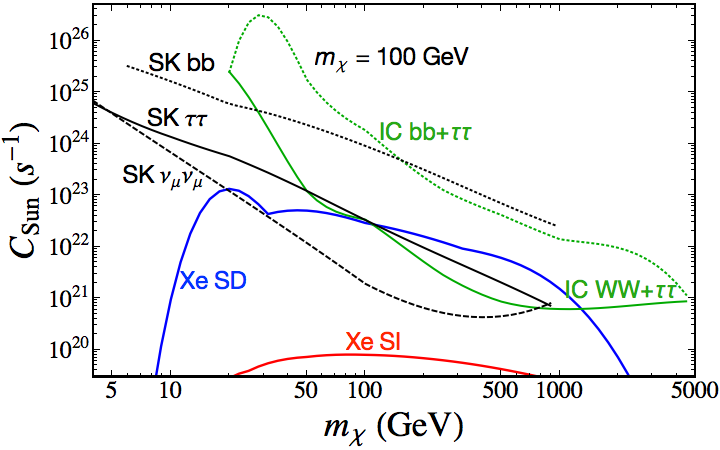}~\includegraphics[width=0.49\textwidth]{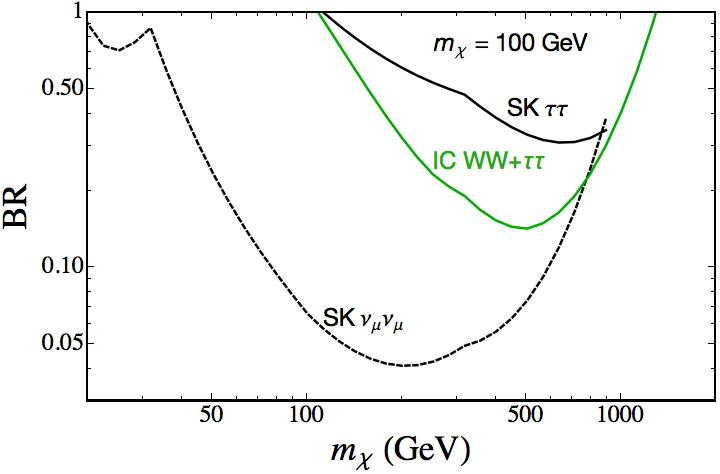}\\
	\includegraphics[width=0.5\textwidth]{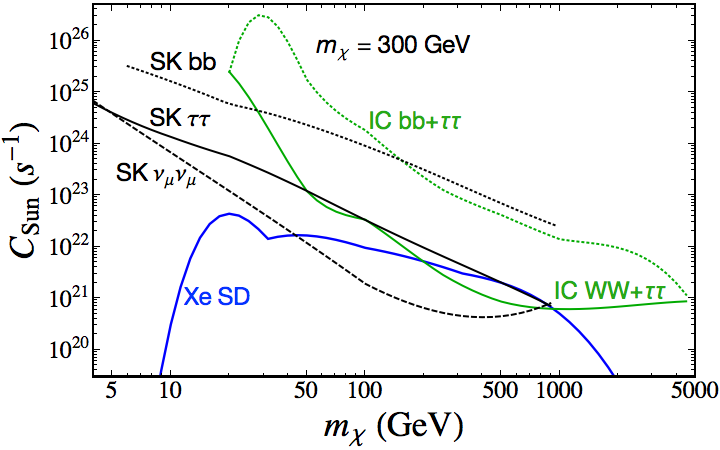}~\includegraphics[width=0.49\textwidth]{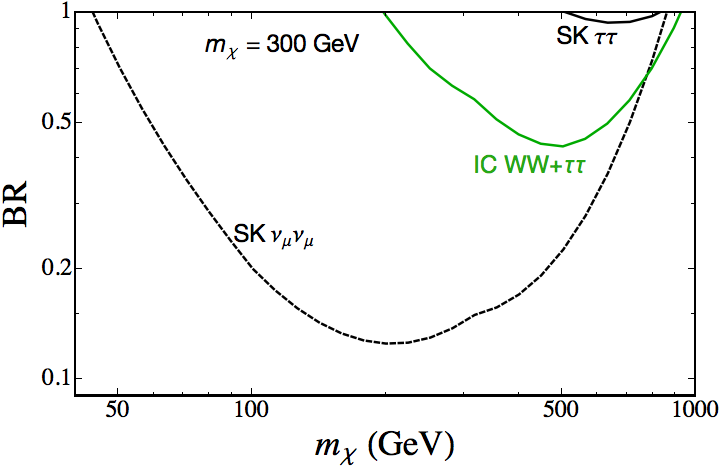}
  \caption{Left: Lower bounds on the capture rate from a future xenon
    DD experiment for SI (red) and SD (blue) interactions compared to
    the $90\%$~CL upper limits for the combined annihilation channels
    $WW,\tau\tau$ and $bb,\tau\tau$ from IceCube (IC, green curves)
    \cite{Aartsen:2012kia} and the channels $\nu_\mu \nu_\mu$,
    $\tau\tau$, $bb$ from Super-Kamiokande (SK, black curves)
    \cite{Tanaka:2011uf, Guo:2013ypa, Choi:2015ara}. Right: Upper bounds on the
    branching ratios versus dark matter mass for SD interactions.  To
    calculate the DD mock data we assume a ``true value'' for the DM
    mass of $m_\chi=30, \, 100,\, 300$~GeV in the top, middle, bottom
    panels, respectively, and values of $10^{-45}\, \rm cm^2$ for the
    SI cross section and $2\cdot 10^{-40}\, \rm cm^2$ for the SD cross
    section, with equal couplings to protons and neutrons in both
    cases. Assumptions about the mock data for the DD experiment are
    given in section~\ref{mock}.} \label{XenonboundsDM}
\end{figure}

In figure~\ref{XenonboundsDM} we show the lower bounds on the capture,
assuming a true (but unknown) dark matter mass of $m_{\chi}=30$~GeV
(upper panels), $100$~GeV (middle panels), and $300$~GeV (lower panels) for a xenon DD
experiment and compare them to the limits from IC and SK. While for SI
interactions the lower bound from DD would be consistent with the
limits from IC and SK, we see that for SD interactions tension arises
if DM annihilates into neutrinos or into $\tau \tau/WW$. In the right
panels we show the ratio of the upper limit on $C_{\rm Sun}$ to the
lower bound. This can be interpreted as an upper bound on the
branching ratio of the corresponding annihilation channel. This is
conservative, since it assumes that there are no neutrinos from any
other annihilation channels, which typically will not be the case.

We see that direct annihilations into neutrinos would be constrained
to branching ratios at the few~\% level, with some dependence on the DM
mass. Annihilations into $\tau\tau, WW$ could be
constrained at the 10\% level. However, note that for true $m_\chi
\lesssim 100$~GeV the bound is stronger when applied at a ``wrong''
DM mass, i.e., for values of $m_\chi$ which are larger than the actual
true value. When using the correct value of $m_\chi$ the limits for
the assumptions adopted here would be rather weak.

\begin{figure}[t]
	\centering
	\includegraphics[width=0.5\textwidth]{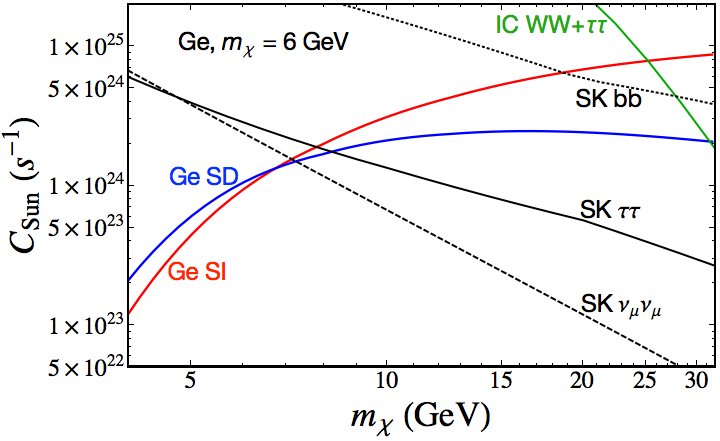}~\includegraphics[width=0.48\textwidth]{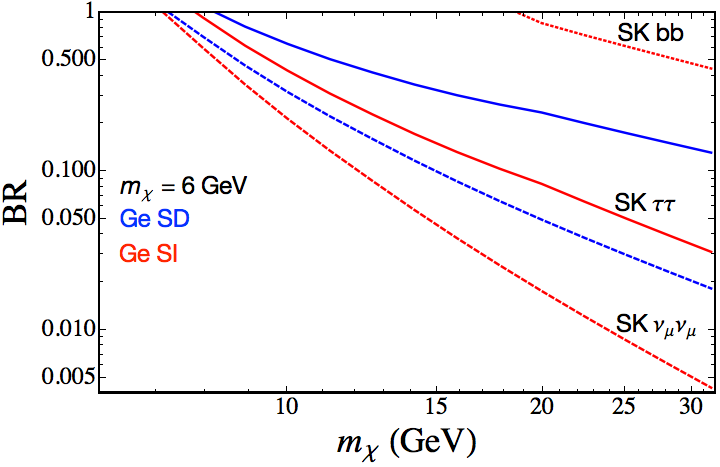}
  \caption{Left: Lower bounds on the capture rate from a future
    germanium DD experiment for SI (red) and SD (blue) interactions
    compared to the $90\%$~CL upper limits for the combined
    annihilation channels $WW,\tau\tau$ from IceCube (IC, green
    curves) \cite{Aartsen:2012kia} and the channels $\nu_\mu \nu_\mu$,
    $\tau\tau$, $bb$ from Super-Kamiokande (SK, black curves)
    \cite{Tanaka:2011uf, Guo:2013ypa, Choi:2015ara}. Right: Upper bounds on the
    branching ratios versus dark matter mass.  To calculate the DD
    mock data we assume a ``true value'' for the DM mass of
    $m_\chi=6$~GeV and cross sections of $\sigma_{\rm SI}=5\cdot
    10^{-42}\, \rm cm^2$ and $\sigma_{\rm SD}=2\cdot 10^{-40}\, \rm
    cm^2$ (equal couplings to protons and neutrons). Assumptions about
    the mock data for the DD experiment are given in
    section~\ref{mock}.}\label{GeboundsDM}
\end{figure}

In figure~\ref{GeboundsDM} we show the lower bounds on the capture
(left) and the upper bounds on the branching ratios for different
channels (right), assuming the low threshold germanium
experiment. Since this configuration is most sensitive at low DM
masses we take here a ``true'' DM mass of 6~GeV. In this case SD and
SI interactions lead to similar lower bounds which potentially can
constrain annihilations into channels leading to neutrinos. Again we
note the feature that bounds get stronger when applied for the ``wrong''
DM mass, in which annihilations into $\tau\tau$ become also constraining.

We note that our comparison of DD and neutrino data is not fully
consistent, in the sense that we are comparing a possible signal in a
future DD experiment with current limits from neutrino
telescopes. When the potential signals from DD will be available,
limits from IC and/or SK may have improved and upgraded and/or new
neutrino telescopes may be in place \cite{Aartsen:2014oha, km3net,
  Abe:2011ts}. In this sense our results are conservative, since the
comparison may become more stringent than what is shown here.

Let us comment briefly on the annual modulation of the DD signal. We
expect that using information on the modulation in addition to the
unmodulated rate does not provide significantly stronger bounds on the
capture rate. If a future experiment only measures the modulation
amplitude but cannot distinguish the unmodulated DM rate from
background (similar to the DAMA experiment \cite{Bernabei:2010mq}) one can use the results of
refs.~\cite{HerreroGarcia:2011aa,HerreroGarcia:2012fu} to derive a
lower bound on $C_{\rm Sum}$ based on the modulation amplitude. Those
bounds require some additional (modest) assumptions on the halo and
they can be found in appendix~\ref{ap:modulations}, where we also
apply those bounds for the DAMA signal \cite{Bernabei:2010mq}, which however is incompatible
with other DD limits halo-independently \cite{HerreroGarcia:2012fu}.

\subsection{Uncertainties due to form factors and unknown couplings to protons and neutrons}
\label{uncertainties}

In this section we discuss the uncertainties associated to nuclear
form factors, as well as the case of arbitrary couplings to neutrons
and protons. Different calculations of form factors sometimes lead to
large differences, especially for SD interactions and heavy nuclei
such as xenon, see for instance refs.~\cite{Bednyakov:2006ux,
  Toivanen:2009zza, Cerdeno:2012ix, Klos:2013rwa,
  Aprile:2013doa} for discussions. In our numerical calculations based
on mock data no form factor appears, since we assume that the
``correct'' one is used, when extracting the velocity distribution
from the observed rate via eq.~\eqref{eq:deriv}. If a ``wrong'' form
factor was used in eq.~\eqref{eq:deriv} it would modify the extracted
velocity distribution, $\tilde f_\text{extr}(v)$, in the following
way:
  \begin{equation}\label{eq:derivFF}
\mathcal{C} \tilde f_\text{extr}(v) =  
\mathcal{C} \tilde f(v) \frac{F^2_\text{true}(E_R)}{F^2_\text{wrong}(E_R)} -
\frac{\tilde\eta(v)}{v} \frac{d}{d v}
\left(\frac{F^2_\text{true}(E_R)}{F^2_\text{wrong}(E_R)} \right) \,.
  \end{equation}
Let us focus on the case of SD interactions (where form factor
uncertainties can be large) and consider also general couplings to
neutrons ($a_n$) and protons ($a_p$). In this case the form factor can be
expressed in terms of the structure functions $S_{ij}(E_R)$ as
\begin{equation}\label{eq:FFDF}
F_{\rm SD}^2(E_R) = \Big(1+\kappa \Big)^2\,S_{00}(E_R)+\Big(1-\kappa^2\Big)\,S_{01}(E_R)+\Big(1-\kappa\Big)^2\,S_{11}(E_R) \,,
\end{equation}
where we define $\kappa\equiv a_n/a_p$ and we have absorbed $a_p$ in
the cross section $\sigma_\chi^p$.  Now we can use this relation
together with eq.~\eqref{eq:derivFF} to test both the impact of using
a ``wrong'' form factor as well as ``wrong'' values of $\kappa$ when
deriving the bound on the capture rate from mock data.

In order to illustrate those effects we generate mock data assuming SD
interactions for a neutron-dominated experiment (xenon) and for a
proton-dominated one (fluorine). We use the same parameters for the
mock data as in section~\ref{mock} for both experiments, with
$\sigma_{\rm SD}=2\cdot 10^{-40}\, \rm cm^2$, a DM mass of
$m_\chi=100$~GeV, and $E_{\rm thr}=3$~keV. For these parameter values,
we predict $78$ events in the energy range $[3,10]$~keV for a fluorine
experiment with total exposure of $100\, \rm kg\cdot y$.
To generate mock data we assume equal couplings to protons and
neutrons ($\kappa=1$).

\begin{figure}[t]
	\centering
        \includegraphics[width=0.62\textwidth]{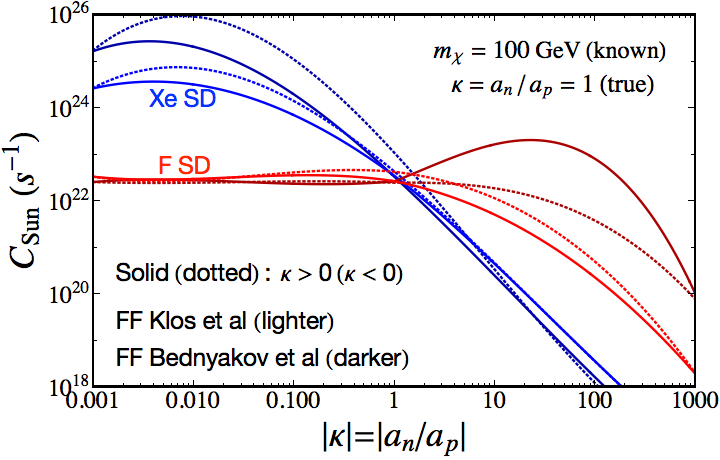}
	\caption{Lower bound on the capture rate for SD interactions
          from mock data from experiments using xenon (red) and
          fluorine (blue). Mock data is generated for a DM mass of
          $m_\chi=100$~GeV, $\sigma_\chi^p =2\cdot 10^{-40}\, \rm
          cm^2$, and equal couplings to protons and neutrons ($\kappa=
          a_n / a_p = 1$). The lower bound on $C_{\rm Sun}$ is shown
          as a function of $|\kappa|$ (i.e., using a ``wrong'' value
          of $\kappa$ when using the mock data to calculate the
          bound), where the solid (dotted) curves correspond to
          $\kappa > 0$ ($\kappa < 0$). Dark (light) colours correspond to the form factors from 
          ref.~\cite{Bednyakov:2006ux} (ref.~\cite{Klos:2013rwa}).} \label{capanap}
\end{figure}

In figure~\ref{capanap} we show the lower bound on the capture rate as
a function of $\kappa$, illustrating the case of using a ``wrong''
ratio of neutron to proton couplings when analysing the data. For $\kappa \sim1$ both F and Xe give similar lower bounds, with the difference coming from their minimum velocities at the threshold ($v_{\rm thr}$) due to their different masses. For the xenon experiment, for $|\kappa| \gg 1$, the capture rate vanishes,
since in xenon spin is mostly carried by the neutron. Hence, for
$|\kappa| \gg 1$ the DD signal is dominated completely by neutrons and
since the capture in the Sun is set by protons, $C_{\rm Sun}$ becomes
suppressed by $1/\kappa^2$ in that case. On the other hand, the spin
of the fluorine nucleus is mainly provided by a proton. Hence, for
$|\kappa| < 1$ the bound on the capture rate becomes independent of
the ratio of the couplings, since both DD and $C_{\rm Sun}$ are
controlled by interactions with protons.

We also illustrate the effect of using different form factors in
figure~\ref{capanap}, comparing form factor calculations from
ref.~\cite{Bednyakov:2006ux} (darker colours) and
ref.~\cite{Klos:2013rwa} (lighter colours). We see that those two
examples give similar results in the regime where the interaction is
``large'', i.e., for neutron (proton) dominated interactions for xenon
(fluorine). However, in the opposite case those two form factor
calculations lead to very different results. For $|\kappa| < 1$
(proton domination), a xenon experiment would give very strong bounds
on the capture rate, since interactions are suppressed for DD, and
explaining the assumed signal would require a very large scattering cross
section. However, in this regime the bounds differ by about one order
of magnitude between the two form factor models. Similarly, for
$|\kappa|> 1$ (neutron domination), the bound on the capture rate from
a fluorine experiment changes by more than a factor 100 between the
two form factor calculations. 

Let us stress an important point related to using ``wrong'' values of
$\kappa$. This discussion is mostly relevant if data is available only
from one experiment (or experiments with the same spin structure). If
a significant signal from xenon as well as fluorine is observed (such
as assumed in our mock data), then the neutron to proton ratio
$\kappa$ is essentially determined by the relative strength of the two
signals. This emphasizes the need of data from complementary targets.
A similar discussion will also apply for SI interactions with
arbitrary couplings to neutrons and protons.  We leave a detailed
study of SI interactions with general isospin structure for future
work.

In figure~\ref{capanap} we have changed
the form factors both for generating mock data as well as calculating
the bound on the capture rate, i.e., we have always used the
``correct'' form factor.  We have also tested the impact of adopting a
``wrong'' form factor by using eq.~\eqref{eq:derivFF}.  By comparing
the form factors from
refs.~\cite{Bednyakov:2006ux} and \cite{Klos:2013rwa} in this way we
find in some cases even unphysical negative values for $\tilde
f_\text{extr}(v)$. This means that the wrong form factor leads to
inconsistent results and one would be able to see from the observed
spectrum that data are not consistent with that particular form factor
choice. Hence, we note that form factor uncertainties are important,
but once a precise spectrum from DD is available (such as necessary
for our method to work) we will have an additional tool at hand to
test different form factor models.

\section{Discussion and conclusions} \label{conc}

We have established a halo-independent framework to relate a signal in
a DM direct detection experiment to the neutrino rate in neutrino
telescopes from DM annihilations in the Sun. Assuming that the DM
velocity distribution and the DM density are constant on time scales
relevant for equilibration in the Sun, we have derived a lower bound
for the DM capture rate in the Sun in terms of a positive signal in a
direct detection experiment, see eq.~\eqref{SI}. If DM capture and
annihilation in the Sun are in equilibrium we obtain (conservative)
upper bounds on branching fractions for annihilations in channels
involving neutrinos from the comparison of the lower bound on the
capture rate from a direct detection signal with the upper limits from
neutrino telescopes.

The lower bounds are based on the part in DM velocity space which
contributes to both the capture in the Sun as well as the scattering
in direct detection experiments. We find that for typical threshold
energies of direct detection experiments a significant overlap region
exists to apply our bounds. Hence, the lower bounds are independent of
the velocity distribution, the escape velocity, the cross section or
the local DM density, although we implicitly assume that these values
are such that a direct detection signal can be measured. With some
additional modest assumptions on the halo properties the bound can
also be used for an annual modulation signal, as explained in
appendix~\ref{ap:modulations}.

To illustrate the power of the bounds, we have applied them to mock
data from future xenon and germanium experiments, assuming that the
true DM cross section is not too far from the current upper bounds. In
such a case the halo-independent comparison to present limits from
Super-Kamiokande and IceCube leads to non-trivial bounds on the
branching fraction for direct annihilations into neutrinos, if the
scattering is spin-dependent.  In some cases also annihilations into
$\tau\tau$ start getting constrained. For spin-independent
interactions current constraints are weak. In general we note that
bounds are stronger for DM masses in the range between 100 to
500~GeV. For lower DM masses the lower bound from direct detection as
well as the limits from neutrino telescopes become weaker. Those
results can be found in figures~\ref{XenonboundsDM} and
\ref{GeboundsDM}. We note that this method is expected to become more
powerful when data from future neutrino telescopes such as
IceCube-Pingu, Hyper-Kamiokande, or KM3NET become available.

The halo-independent comparison has to be done for a specific model
for the DM--nucleon interaction (and as a function of the DM mass).
In our work we mostly assumed either SI or SD elastic
scattering with equal couplings to neutrons and
protons. Generalization to other types of couplings is
straightforward. As an example we discussed in
  section~\ref{uncertainties} the case of SD interactions with
  arbitrary couplings to neutrons and protons.  This leads to
  interesting situations, since only scattering on free protons is
relevant for the capture in the Sun, whereas for direct detection the
sensitivity is governed by the spin-composition of the target
nucleus. One may imagine a situation where the spin of the nucleus is
dominated by neutrons, in which case the neutrino and direct detection
signals largely decouple. In order to exclude such a case it will be
essential that data from target nuclei with spin carried by protons is
available. Another case in which a DD signal does not
  always imply a neutrino signal from the Sun is asymmetric DM (see ref.~\cite{Zurek:2013wia} for a recent review). While
  our bounds on the capture rate would still apply, in those models
  annihilations are often suppressed due to the lack of anti-DM, and therefore a possible signal from
  neutrinos becomes very model dependent.

In our work we always assumed that the
differential scattering cross section $d\sigma/dE_R$ is proportional
to $1/v^2$, which is true for DM--nucleon contact interactions, but in
general this may not be the case for more exotic types of interactions
(see for instance ref.~\cite{DelNobile:2013cva} for generalizing
halo-independent methods for direct detection to such cases).  In
appendix~\ref{app:v} we show that it is straightforward to generalize
our lower bound on the capture rate to all cases where the velocity and
nuclear recoil energy dependence of the differential cross section
factorizes as $d\sigma/dE_R = g(v) h(E_R)$. Considering more
complicated models is beyond the scope of this paper and is left for
future work. Another particle physics variation for which our results
do not apply directly is inelastic scattering of the type $\chi + N
\to \chi^* + N$, where the mass difference between $\chi$ and $\chi^*$
is of the order of the DM kinetic energy. This will change the
kinematics of the scattering and generically for $m_{\chi^*} > m_\chi$
the capture rate is increased \cite{Nussinov:2009ft, Menon:2009qj, Shu:2010ta},
see ref.~\cite{Bozorgnia:2013hsa} for halo-independent considerations in
the context of DD. We leave also the generalization of the lower bound
on the capture rate to this case for future work.

To conclude, we would like to emphasize that in the
  presence of a DD signal and a signal of neutrinos from the Sun,
  before proceeding to extract the parameters by doing a fit to both,
  one should first check if the lower bounds on the capture derived
  here are fulfilled for some combination of annihilation channels and
  branching ratios. This would provide a halo-independent consistency
  check for both signals.

\bigskip

{\bf Acknowledgements:} We thank Joakim Edsj\"o for useful discussions. This work was supported by the G\"oran Gustafsson Foundation [M.B.]. We also thank the Nordita Scientific Program ``News in Neutrino Physics'', where this work was initiated.

\appendix
\section{Lower bound on the capture for annual modulation signals} 
\label{ap:modulations}

The change of the velocity of the detector relative to the DM halo due
to the Earth's rotation around the Sun leads to an annual modulation
of the event rate in a DD experiment \cite{Drukier:1986tm,
  Freese:1987wu}. We denote the amplitude of the modulation expressed
in $v_m$ space by $A_\eta(v_m)$.  Information from a modulation signal
can be combined with the halo-independent lower bound on the capture
rate, eq.~\eqref{SI}, by using halo-independent upper bounds on the
annual modulation in terms of the average rate
$\overline{\eta}(v_m)$ derived in ref.~\cite{HerreroGarcia:2011aa} and
applied in refs.~\cite{HerreroGarcia:2012fu,Bozorgnia:2013hsa}. Those bounds
 are based on an
expansion in the Earth velocity $v_e$, using that $v_e/v$ is small for
$v \ge v_m$ for typical values of $v_m$ relevant for experiments. In
such an expansion the time independent rate appears at zeroth order,
whereas the annual modulation amplitude is of linear order in $v_e$.

If we assume that the DM velocity distribution is constant on time scales of years and constant in space on scales of the Sun-Earth distance, one can derive the bound \cite{HerreroGarcia:2011aa}
\beq \label{eq:bound_gen}
A_\eta(v_m) \leq v_e 
\left[-\frac{d \overline \eta}{d v_m}+ 
    \frac{\overline{\eta}(v_m)}{v_m}  
   - \int_{v_m} dv \frac{\overline{\eta}(v)}{v^2} 
\right] \,.
\eeq
Note however, that the lower bounds on the capture rate in
eq.~\eqref{SI} require a lower bound either on $(-d\overline\eta/dv_m)$
or on $\overline\eta(v_m)$, which cannot be obtained in terms of
$A_\eta(v_m)$ alone from eq.~\eqref{eq:bound_gen}. In principle
eq.~\eqref{eq:bound_gen} can be re-written as a lower bound on
$(-d\overline\eta/dv_m)$ involving both, $A_\eta(v_m)$ and
$\overline\eta(v_m)$. Hence, this would require an experiment able to
determine the modulation amplitude as well as the unmodulated
rate. Moreover we do not expect a significant stronger lower bound on
$C_{\rm Sun}$ from such a procedure for modulation amplitudes which
fulfill eq.~\eqref{eq:bound_gen} (as they must to be consistent with a DM signal).

Adopting some modest additional assumptions on the halo another bound
on the modulation amplitude can be derived. If there is only one
preferred direction in the DM velocity distribution one finds (see ref.~\cite{HerreroGarcia:2011aa} for details):
\beq \label{eq:bound_sym}
A_\eta(v_m) \leq - v_e \sin\alpha_{\rm halo}
\frac{d \overline \eta}{d v_m} ,
\eeq
where $\alpha$ is the angle between the preferred DM direction and the
direction perpendicular to the Earth's orbit. Assuming an observed signal for $A_\eta(v_m)$,
this provides us with a lower bound on 
$(-d\overline\eta/dv_m)$, which can be directly plugged in eq.~\eqref{SI} and we get:
\beq  \label{Cboundmodb}
C_{\rm Sun} \geq 4\pi \sum_A A^2 \int_0^{R_{\rm Sun}} dr r^2 \rho_A(r) \int_{v_{\rm thr}}^{v_{\rm cross}^A} dv\, \frac{\tilde{A}_\eta(v)}{\sin\alpha_{\rm halo}\,v_e}\, \mathcal{F}_A(v),
\eeq
where in the last line we used $\tilde{A}_\eta(v)\equiv
\mathcal{C}\,A_\eta(v)$. In the following we will adopt the most
conservative option $\sin\alpha_{\rm halo} \simeq 1$, but notice that in
cases where the preferred DM direction is set by the direction of the
Sun relative to the DM halo (e.g., for a halo with a dark disk) one
has $\sin\alpha_{\rm halo}\simeq 0.5$ and the bounds scale accordingly.

Notice that observing annual modulation with typical exposures (and
allowed cross section values) is extremely hard (see for instance
ref.~\cite{Bozorgnia:2014dqa}). The DAMA/LIBRA experiment reports an annual
modulation of the signal in their NaI scintillator detector, with a
period of one year and a maximum around June 2nd with very high
statistical significance \cite{Bernabei:2010mq}. This modulation is
strongly disfavoured by other experiments halo-independently, both for
elastic SI and SD interactions \cite{HerreroGarcia:2012fu} and for
inelastic scattering \cite{Bozorgnia:2013hsa}. Despite these problems of the
DM interpretation of the DAMA modulation signal we will use it in the
following to illustrate the lower bounds on the capture rate
eq.~\eqref{Cboundmodb} based on a modulation signal.

\subsection{The DAMA modulation signal}

\begin{figure}[t]
	\centering
\includegraphics[width=0.5\textwidth]{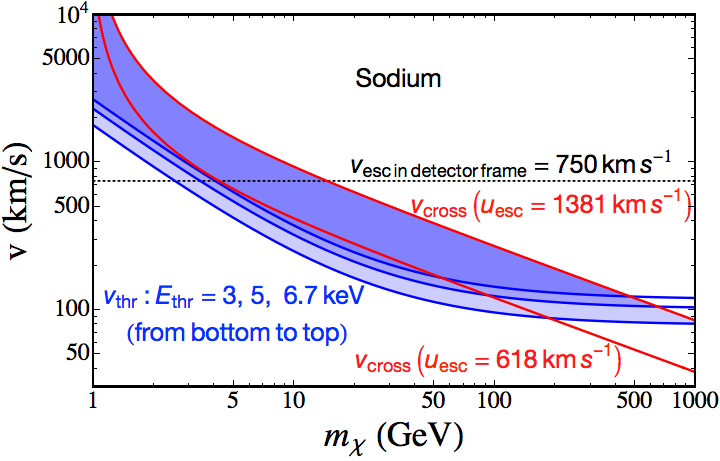}~\includegraphics[width=0.5\textwidth]{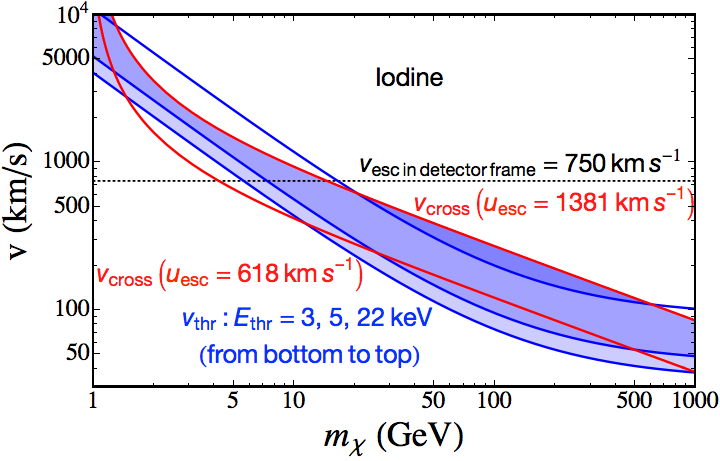}
  \caption{In blue we show the minimum velocity $v_{\rm thr}(E_{\rm
      thr})$ probed in a direct detection experiment versus $m_\chi$,
    assuming different threshold energies $E_{\rm thr}$ and using a Na
    (left) and a I (right) target. In red we show the maximum
    velocity relevant for DM capture in the Sun, $v_{\rm cross}^p$ for
    scattering on hydrogen for the two extreme values of $u_{\rm esc}
    = 1381 \, {\rm km \, s^{-1}}$ in the centre of the Sun and $v_{\rm
      esc} = 618 \, {\rm km \, s^{-1}}$ at the surface. The
    shaded area shows the overlap region assuming scattering in the
    centre of the Sun. The horizontal
    black line indicates approximately the galactic escape velocity in
    the detector rest frame.} \label{DIDDAMA}
\end{figure}

In figure~\ref{DIDDAMA} we show the overlap in $v_m$ space between
DAMA and the DM capture in the Sun assuming scattering either on Na or
on I. Na dominates for DM masses $m_\chi \lesssim 20$~GeV, whereas
iodine is relevant for larger DM masses. For sodium we observe a large
overlap region. Note that the DAMA threshold of 2~keVee corresponds to
a recoil threshold of 6.7~keV (Na) and 22~keV (I) for usual quenching
factors. This implies a relatively small overlap region in the case of
iodine, especially for the SD case shown in figure~\ref{DIDDAMA}.

\begin{figure}[t]
\centering
\includegraphics[width=0.5\textwidth]{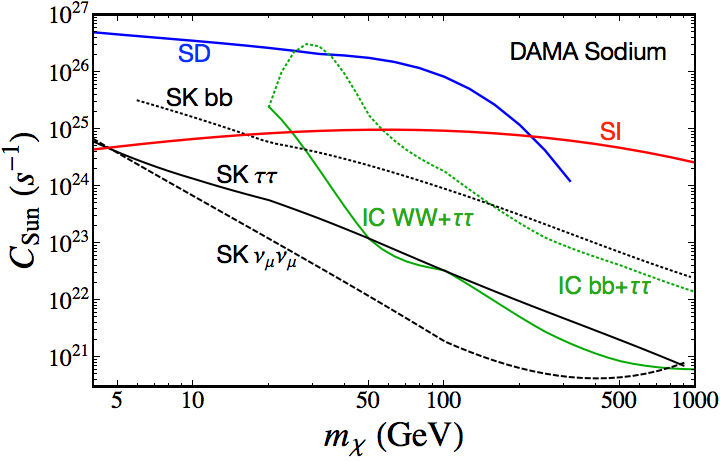}~\includegraphics[width=0.5\textwidth]{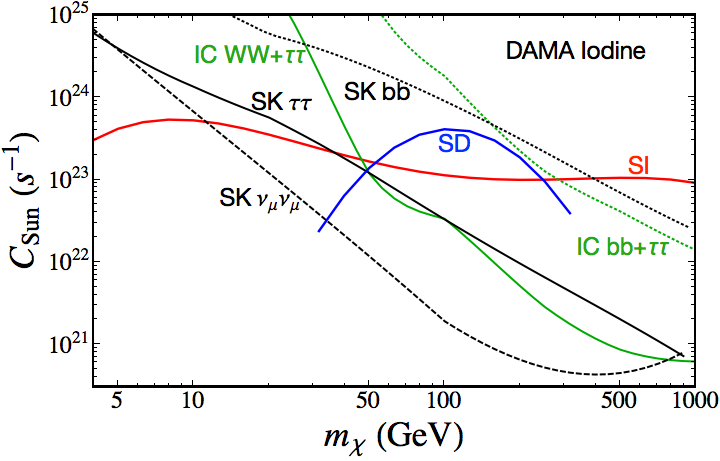}
\caption{Lower bound on the capture rate from DAMA data assuming
  scattering on sodium (left) or iodine (right) for SI (red) and SD
  (blue) interactions, using the bound from
  eq.~\eqref{Cboundmodb}. Also shown are the $90\%$~CL upper bounds
  from IceCube \cite{Aartsen:2012kia} for annihilations into
  $WW+\tau\tau$ and $WW+bb$ (green curves), and from Super-Kamiokande
  into $bb$, $\tau\tau$, and $\nu_\mu\nu_\mu$ \cite{Tanaka:2011uf,
    Guo:2013ypa, Choi:2015ara} (black curves).} \label{DAMAbounds}
\end{figure}

To obtain a lower bound on the capture rate from DAMA data we consider
a binned version of eq.~\eqref{Cboundmodb} and use the observed values
of $\tilde A_\eta(v_m)$ corresponding to the energy bins reported by
DAMA, see refs.~\cite{HerreroGarcia:2011aa, HerreroGarcia:2012fu,
  Bozorgnia:2013hsa} for details on this procedure.  We assume either
elastic SI or SD scattering on Na or on I with equal couplings to
protons and neutrons.  The resulting lower bounds on the capture rate
are shown in figure~\ref{DAMAbounds} both for SI and SD, together with
the $90\%$~CL upper limits on the capture from neutrino telescopes.

From the left panel we see that for scattering on sodium there is
tension between the lower and upper bounds implying that for SI DM
annihilation into neutrinos and $\tau\tau$ ($bb$) are strongly constrained
for $m_\chi \gtrsim 5$~GeV ($15$~GeV), while SD is excluded for all channels.

For scattering on iodine
shown in the right panel of figure~\ref{DAMAbounds} we note the
strong dependence on the DM mass for SD scattering. This can be
understood from figure~\ref{DIDDAMA}, which shows that only for a
small range of DM masses there is overlap in $v_m$ space. For SI
interactions $v_{\rm cross}$ is larger (not shown in the plot),
leading to larger overlap in $v_m$ space, and we observe strong bounds
on the annihilations shown in figure~\ref{DAMAbounds} for $m_\chi
\gtrsim 35$~GeV (and even for $m_\chi \gtrsim 10$~GeV for the neutrino
channel).

\section{Beyond contact interactions}
\label{app:v}

In the main text of the paper we have always assumed contact
interactions between DM and the nucleus, which leads to a $1/v^2$
dependence of the differential scattering cross section. In more
exotic models also other dependences on $v$ and/or $E_R$ are possible,
see for instance refs.~\cite{DelNobile:2013cva, Guo:2013ypa, Liang:2013dsa} in the context of DD and the
neutrino signal. Our bounds can be generalized in a straightforward
way if the dependence of $v$ and $E_R$ factorizes. Let us assume that
the differential cross section on the nucleus with mass number $A$ can
be written in the form
\begin{align}\label{CS-nonst}
 \frac{d\sigma_A}{dE_R} = g_A(v) h_A(E_R) \,. 
\end{align}
In the conventional case considered in the main text we have $g_A(v)
\propto 1/v^2$ and $h_A(E_R) = F_A^2(E_R)$, see eq.~\eqref{eq:CS}.

From the measured recoil spectrum in a DD experiment,
$\mathcal{R}(E_R)$, we can then extract the DM velocity distribution by
\begin{align}
  \tilde f(v)\,v = - \frac{m_\chi m_A}{\rho_\chi v^2 g_A(v)} \frac{d}{dv} 
  \left(\frac{\mathcal{R}(E_R)}{h_A(E_R)}\right) \,.
\end{align}
Using this in the expression for the capture rate in the Sun (eqs.~\eqref{capt} and ~\eqref{Omega}) we find
that the bound eq.~\eqref{SI} becomes
\begin{align}\label{bound-nonst}
  C_{\rm Sun} \ge 4\pi \sum_A \int_0^{R_{\rm Sun}} dr r^2 \rho_A(r)
  \int_{v_{\rm thr}}^{v_{\rm cross}^A} dv 
  \left[- \frac{d}{dv} \left(\frac{\mathcal{R}(E_R)}{h_{A_{\rm DD}}(E_R)}\right) \right]
  \frac{w^2 g_A(w)}{v^2 g_{A_{\rm DD}}(v)} \mathcal{H}_A(E_R, r) 
\end{align}
where $A_{\rm DD}$ indicates the mass number of the nucleus in the
DD experiment, whereas the sum over $A$ runs over the elements in the
Sun, $w^2 = v^2 + u_{\rm esc}^2(r)$, and
\begin{align}
    \mathcal{H}_A(E_R, r) \equiv \int_{E_{\rm min}(v)}^{E_{\rm max}(v)} h_A(E_R) dE_R \,.
\end{align}
Eq.~\eqref{bound-nonst} corresponds to the lower bound on the
capture rate if the scattering cross section can be factorized
according to eq.~\eqref{CS-nonst}. For certain models such a
factorization may not be possible. Generalizing our bound to those
cases is beyond the scope of this paper and we leave it for future work.

\bibliographystyle{my-h-physrev.bst}
\bibliography{paper}

\end{document}